\documentstyle[preprint,aps]{revtex}
\begin{document}
\title{''All-versus-nothing'' nonlocality test of quantum mechanics by two-photon
hyperentanglement}
\author{C. Cinelli, M. Barbieri, R. Perris, P. Mataloni and F. De Martini.}
\address{Dipartimento di Fisica dell' Universita' ''La Sapienza'' and Consorzio\\
Nazionale Interuniversitario per\\
le Scienze Fisiche della Materia, Roma, 00185 Italy}
\maketitle

\begin{abstract}
We report the experimental realization and the characterization of
polarization and momentum hyperentangled two photon states, generated by a
new parametric source of correlated photon pairs. By adoption of these
states an ''all versus nothing'' test of quantum mechanics was performed.
The two photon hyperentangled states are expected to find at an increasing
rate a widespread application in state engineering and quantum information.

\end{abstract}


The violation of Bell's inequalities has been recognized as the first,
paradigmatic test of quantum nonlocality \cite{1,2}. As such, in the last
decades it has been somewhat successfully realized by many two-particle
experiments mostly performed by optical techniques \cite{3,4}. However, in
spite of the fact that today bipartite pure two-photon
polarization-entangled states can be rather easily produced by Spontaneous
Parametric Down Conversion (SPDC) in a non linear (NL) crystal, a quite
unsatisfactory feature of the Bell's inequality method is that its probative
effectiveness only applies to statistical measurement procedures. In the
framework of this method the EPR\ local realistic picture can indeed explain
perfect correlations implied by predictions to be tested by any definite,
single experiment. This shortcoming does not affect the Hardy's
''nonlocality test without inequalities'', also referred to as ''ladder
proof of nonlocality'', where a contradiction with Einstein's local realism
can be demonstrated, at least in principle, by any single experimental test
but only for a small fraction of the generated photon pairs \cite{5,6}. The
extension to the full set of pairs is important as it provides a complete
and startling demonstration of the conflict existing between the laws of
quantum mechanics and the expectations of local realism. This test is
offered by the ''all versus nothing'' nonlocality proof \cite{7} and is
based on the Grrenberg-Horne-Zeilinger (GHZ) theorem \cite{8}. It applies
for the set of systems that are in the same GHZ state and has been
experimentally demonstrated either for three- or four-photon entanglement 
\cite{9,10,11}. In addition to their conceptual relevance, tests of quantum
mechanics performed by states operating in a large dimension Hilbert space
exhibit deviations from local realist expectations which are larger and more
robust against noise \cite{12}.

Recently, it has been suggested that the realization of GHZ theorem can be
extended to the case of two photon hyper-entangled states in a $(4x4)$
Hilbert space \cite{13}, i.e. simultaneously maximally entangled in two
degrees of freedom, as for instance the field's polarization $(\pi )$ and
the spatial momentum $({\bf k)}$ \cite{14}. In this way the intrinsic
limitation of SPDC where no more than one photon pair is created in any
elementary annihilation-creation process can be easily overcome. In most
protocols of quantum information (QI) these $2-$fold hyper-entangled states
associated to a pair of correlated particles act as $4-$particles in the
usual entanglement configuration. Indeed, the first and foremost application
of the $(\pi -{\bf k)}$ entanglement was the first Quantum State
Teleportation experiment carried out in Rome in 1997 \cite{15}. In that
experiment the complete analysis at the Alice's site of the four orthogonal
Bell states with $100\%$ efficiency was realized, a result otherwise
impossible to achieve with standard entanglement and linear optical\
techniques \cite{16}. In this letter we report on the systematic production
and characterization of $(\pi -{\bf k)}$ hyper-entangled photon pairs,
generated by an ''ad hoc'' flexible parametric source recently developed in
our laboratory. This device will be then adopted to realize the experimental
''all versus nothing'' test of quantum nonlocality.

The source adopted in this experiment is schematically shown in Fig. 1a. It
has been extensively described in previous papers \cite{17}. The $2-$photon
states generated over the emission cone of a $0.5$ mm thick $\beta $%
-barium-borate (BBO) type I crystal, were simultaneously entangled in
polarization and momentum. Polarization $(\pi )$ entanglement was obtained
by quantum superposition of the two overlapping radiation cones generated at
the same wavelength $\lambda =728nm$\ by BBO, excited in two opposite
directions by a cw Argon laser at $(\lambda _{p}=\lambda /2)$. The two cones
were then carefully overlapped by means of a spherical back-mirror $(M)$.
The overall radiation is expressed by the maximally entangled state $|\Phi
\rangle =2^{-%
%TCIMACRO{\UNICODE[m]{0xbd}}%
%BeginExpansion
{\frac12}%
%EndExpansion
}\left( |H_{1}\rangle |H_{2}\rangle +e^{i\theta }|V_{1}\rangle |V_{2}\rangle
\right) $ in the horizontal $(H)$ and vertical $(V)$\ polarization basis,
with phase $\theta $\ easily and reliably controlled by micrometric
displacements of the back mirror. It may be locally transformed into the
state $|\Psi \rangle =2^{-%
%TCIMACRO{\UNICODE[m]{0xbd}}%
%BeginExpansion
{\frac12}%
%EndExpansion
}\left( |H_{1}\rangle |V_{2}\rangle +e^{i\theta }|V_{1}\rangle |H_{2}\rangle
\right) $\thinspace by the zero-order $\lambda /2$ waveplate ($HW^{\ast }$)
inserted in one of the two correlated directions: Fig. 2a. Entangled states,
either pure or controllable mixed states, have been created in a flexible
way by the same source \cite{18}. Momentum $({\bf k})$ entanglement was
realized, under excitation of either one of the two overlapped radiation
cones, by a four hole screen which allowed to select two pairs of correlated 
$k$-modes, $a_{1}-b_{2}$ and $a_{2}-b_{1}$ within the conical emission of
the crystal: Fig. 1a \cite{19}. The straight lines connecting on the screen
the holes leaving through the correlated pairs intercross at an angle $%
\alpha =18%
%TCIMACRO{\UNICODE[m]{0xb0}}%
%BeginExpansion
{{}^\circ}%
%EndExpansion
$. The ''phase-preserving'' character of the SPDC process allowed to keep
the phase difference $\phi $ between the two pair emission to the value $%
\phi =0$, regardless the value of $\alpha $. The phase $\phi $ could be set
by means of a tilted thin glass plate intercepting mode $a_{2}$. Hence, for
each SPDC generated cone, the ${\bf k}-$ entangled states $|\psi \rangle
=2^{-%
%TCIMACRO{\UNICODE[m]{0xbd}}%
%BeginExpansion
{\frac12}%
%EndExpansion
}\left( |a_{1}\rangle |b_{2}\rangle +e^{i\phi }|b_{1}\rangle |a_{2}\rangle
\right) $ could be generated.

The mode sets $a_{1}-b_{1}$ and $a_{2}-b_{2}$ were split along the vertical
direction by a prism-like two-mirror system and then recombined onto a
symmetric beam splitter $(BS)$: Fig. 2b. A trombone mirror assembly mounted
on a motorized translation stage (not shown in the Figure) allowed fine
adjustments of the path offset delay $\Delta x$ between the input mode pairs 
$a_{1}-b_{1}$ and $a_{2}-b_{2}$. Fig. 1b shows the spatial recombination of
modes $a_{1}-a_{2}$ and $b_{1}-b_{2}$ in two different points of the $BS$
plane.\ The photons associated with the output $BS$ modes, $a_{1}^{\prime
},b_{1}^{\prime }$ and $a_{2}^{\prime },b_{2}^{\prime }$, were independently
detected by four avalanche single photon detectors $(D)$, mod. SPCM-AQR14 $%
D_{a1}$, $D_{b1}$, $D_{a2}$ and $D_{b2}$ in Fig. 1b. This is obtained by
inserting a mirror on each output arm ($M_{1}\,$and $M_{2}$ in Fig 1b),
intercepting the modes $b_{1}^{\prime }$ and $b_{2}^{\prime }$,
respectively. This configuration could be further improved by coupling the
four spatial modes to single mode optical fibers. Equal interference
filters, with bandwidth $\Delta \lambda =6nm$, placed in front of each $D$,
determined the coherence-time of the detected pulses:\ $\tau _{coh}${\it \ }$%
\approx 150f\sec $. Two-photon coincidences were registered for either one
of the following mode combinations: $a_{1}^{\prime }-b_{1}^{\prime }$, $%
a_{1}^{\prime }-b_{2}^{\prime }$, $a_{2}^{\prime }-b_{2}^{\prime }$, $%
a_{2}^{\prime }-b_{1}^{\prime }$, while no coincidence was detected for
modes $a_{1}^{\prime }-a_{2}^{\prime }$ and $b_{1}^{\prime }-b_{2}^{\prime }$%
. The $(\pi -{\bf k)}$ hyper-entangled states realized in the present
experiment could be then expressed as: 
\begin{equation}
\left| \Xi ^{\pm \;\pm }\right\rangle =\left| \Psi ^{\pm }\right\rangle
\otimes \left| \psi ^{\pm }\right\rangle =\frac{1}{2}\left( |H\rangle
|V\rangle \pm |V\rangle |H\rangle \right) \otimes \left( |a_{1}\rangle
|b_{2}\rangle \pm |b_{1}\rangle |a_{2}\rangle \right) \text{,}  \label{hyper}
\end{equation}
where, in $\left| \Psi ^{\pm }\right\rangle $, $\theta =0,\pi $\ and in $%
\left| \psi ^{\pm }\right\rangle $, $\varphi =0,\pi $.

Fig. 1c shows the characteristic quantum resonance effect arising in the $BS$
linear superposition of the two components of any bipartite entangled state 
\cite{20}. It is expressed here for ${\bf k}-$entanglement by the
coincidence rate $C(a_{1}^{\prime },b_{1}^{\prime })$ as a function of $%
\Delta x$. The transition from the symmetric ($\left| \Xi ^{++}\right\rangle 
$) to the antisymmetric ($\left| \Xi ^{+-}\right\rangle $)\ state condition
upon change of the phase $\theta $ is shown with a resonance ''visibility'' $%
\approx 0.90$. Similar results are obtained for the other photon pair
coincidences $C(a_{1}^{\prime },b_{2}^{\prime })$, $C(a_{2}^{\prime
},b_{1}^{\prime })$, $C(a_{2}^{\prime },b_{2}^{\prime })$, measured, by
varying either $\theta $\ or $\varphi $.

Eq. (\ref{hyper}) expresses a two photon hyper-entangled state spanning a $%
(4x4)$ Hilbert space which in the ideal case of a perfect pure state allows
the generalization of the GHZ theorem. This argument is purely logical and
doesn't involve inequalities. However inequalities are necessary as a
quantitative test in a real experiment in order to avoid the conceptual
problems associated to the realization of a null experiment. This can be
given by an ''all versus nothing'' violation of local realism \cite{13}.

In our experiment the hyper-entangled state $|\Xi ^{-\;-}\rangle =\left|
\Psi ^{-}\right\rangle \otimes \left| \psi ^{-}\right\rangle $ has been
adopted to perform the nonlocality test. It is based on the estimation of
the operator ${\cal O}=-z_{1}\cdot z_{2}-z_{1}^{\prime }\cdot z_{2}^{\prime
}-x_{1}\cdot x_{2}-x_{1}^{\prime }\cdot x_{2}^{\prime }+z_{1}z_{1}^{\prime
}\cdot z_{2}\cdot z_{2}^{\prime }+x_{1}x_{1}^{\prime }\cdot x_{2}\cdot
x_{2}^{\prime }+z_{1}\cdot x_{1}^{\prime }\cdot z_{2}x_{2}^{\prime
}+x_{1}\cdot z_{1}^{\prime }\cdot x_{2}z_{2}^{\prime }-z_{1}z_{1}^{\prime
}\cdot x_{1}x_{1}^{\prime }\cdot z_{2}x_{2}^{\prime }\cdot
x_{2}z_{2}^{\prime }$, and the condition for violation of local realism is $%
{\cal O}>7$, while the expected value of ${\cal O}$ according to quantum
mechanics is ${\cal O}=9$ \cite{13}.\ It is worth noting that the local
boundary ${\cal O}\leq 7$ is obtained by employing a hidden variable model
which allows correlations between the $(\pi -{\bf k})$ degrees of freedom of
the same photon. This point has been emphasized by Cabello, who also
stressed in his theory that the present nonlocality test involves only two
observers \cite{11,13}. In the above expression the Pauli operators: 
\begin{eqnarray}
z_{i} &=&\sigma _{z_{i}}=|H\rangle \left\langle H\right| -|V\rangle
\left\langle V\right| \text{, \ \ \ }x_{i}=\sigma _{x_{i}}=|H\rangle
\left\langle V\right| +|V\rangle \left\langle H\right| \text{ \ \ \ \ }%
(i=1,2)  \nonumber \\
z_{1}^{\prime } &=&\sigma _{z_{1}}^{\prime }=|a_{1}\rangle \left\langle
a_{1}\right| -|a_{2}\rangle \left\langle a_{2}\right| \text{, \ \ \ }%
x_{1}^{\prime }=\sigma _{x_{1}}^{\prime }=|a_{1}\rangle \left\langle
a_{2}\right| +|a_{2}\rangle \left\langle a_{1}\right|  \nonumber \\
z_{2}^{\prime } &=&\sigma _{z_{2}}^{\prime }=|b_{1}\rangle \left\langle
b_{1}\right| -|b_{2}\rangle \left\langle b_{2}\right| \text{, \ \ \ }%
x_{2}^{\prime }=\sigma _{x_{2}}^{\prime }=|b_{1}\rangle \left\langle
b_{2}\right| +|b_{2}\rangle \left\langle b_{1}\right| .
\end{eqnarray}
allow the state transformations: 
\begin{eqnarray}
z_{1}\cdot z_{2}|\Xi ^{-\;-}\rangle &=&-|\Xi ^{-\;-}\rangle \text{ \ \ \ }%
z_{1}^{\prime }\cdot z_{2}^{\prime }|\Xi ^{-\;-}\rangle =-|\Xi ^{-\;-}\rangle
\nonumber \\
x_{1}\cdot x_{2}|\Xi ^{-\;-}\rangle &=&-|\Xi ^{-\;-}\rangle \text{ \ \ \ }%
x_{1}^{\prime }\cdot x_{2}^{\prime }|\Xi ^{-\;-}\rangle =-|\Xi ^{-\;-}\rangle
\nonumber \\
z_{1}z_{1}^{\prime }\cdot z_{2}\cdot z_{2}^{\prime }|\Xi ^{-\;-}\rangle
&=&|\Xi ^{-\;-}\rangle \text{ \ \ \ \ }x_{1}x_{1}^{\prime }\cdot x_{2}\cdot
x_{2}^{\prime }|\Xi ^{-\;-}\rangle =|\Xi ^{-\;-}\rangle  \nonumber \\
z_{1}\cdot x_{1}^{\prime }\cdot z_{2}x_{2}^{\prime }|\Xi ^{-\;-}\rangle
&=&|\Xi ^{-\;-}\rangle \text{ \ \ \ \ }x_{1}\cdot z_{1}^{\prime }\cdot
x_{2}z_{2}^{\prime }|\Xi ^{-\;-}\rangle =|\Xi ^{-\;-}\rangle  \nonumber \\
z_{1}z_{1}^{\prime }\cdot x_{1}x_{1}^{\prime }\cdot z_{2}x_{2}^{\prime
}\cdot x_{2}z_{2}^{\prime }|\Xi ^{-\;-}\rangle &=&-|\Xi ^{-\;-}\rangle .
\end{eqnarray}
The nine operators which contribute to determine the expected value of the $%
{\cal O}$ must be evaluated in order to measure a violation of local
realism. It can be observed if the minimum value of entanglement visibility
is $7/9$ \cite{13}. The measured visibilities of polarization and momentum
entanglement obtained by our system are suitable on this purpose.

The experimental apparatuses to observe violation are sketched in Fig. 2a,b.
In both cases the polarization analyzers on each detection arm, $%
a_{1}^{\prime }$, $a_{2}^{\prime }$, $b_{1}^{\prime }$, $b_{2}^{\prime }$
allow to perform the polarization measurement in either the $H-V$\ or the $D-%
\overline{D\text{ }}$basis, with $D=2^{-%
%TCIMACRO{\UNICODE[m]{0xbd}}%
%BeginExpansion
{\frac12}%
%EndExpansion
}\left( H+V\right) $ and $\overline{D\text{ }}=2^{-%
%TCIMACRO{\UNICODE[m]{0xbd}}%
%BeginExpansion
{\frac12}%
%EndExpansion
}\left( H-V\right) .$ The two photons generated by the source are sent to
the Alice and Bob sites which can be indifferently chosen because of the
conical emission symmetry of the parametric radiation. In the present
experiment, Alice and Bob perform the measurements by the upper $%
(D_{a1}-D_{a2})$ and lower $(D_{b1}-D_{b2})$ detectors, respectively. By
referring to the optical setup of Fig. 2a, the modes $a_{1}-a_{2}$ and $%
b_{1}-b_{2}$ are sent directly to the Alice and Bob sites and the
corresponding signals are analyzed by a half wave plate ($HW$) and a
polarizing beam splitter ($PBS$) in each arm. By this apparatus we could
measure the four terms $x_{1}\cdot x_{2}$, $z_{1}\cdot z_{2}$, $%
z_{1}^{\prime }\cdot z_{2}^{\prime }$, $z_{1}z_{1}^{\prime }\cdot z_{2}\cdot
z_{2}^{\prime }$, and $x_{1}\cdot z_{1}^{\prime }\cdot x_{2}z_{2}^{\prime }$.

Fig. 2b shows the optical setup for measuring $x_{1}^{\prime }\cdot
x_{2}^{\prime }$, $x_{1}x_{1}^{\prime }\cdot x_{2}\cdot x_{2}^{\prime }$, $%
z_{1}\cdot x_{1}^{\prime }\cdot z_{2}x_{2}^{\prime }$. Before being analyzed
in each arm and coupled to detectors, the two mode sets $a_{1}-b_{1}$ and $%
a_{2}-b_{2}$ are spatially combined onto the $BS$ which performs the
transformation from the $a_{1}-a_{2}$ to the $d-\overline{d\text{ }}$basis,
where $d=2^{-%
%TCIMACRO{\UNICODE[m]{0xbd}}%
%BeginExpansion
{\frac12}%
%EndExpansion
}\left( a_{1}+a_{2}\right) $ and $\overline{d\text{ }}=2^{-%
%TCIMACRO{\UNICODE[m]{0xbd}}%
%BeginExpansion
{\frac12}%
%EndExpansion
}\left( a_{1}-a_{2}\right) $ and, similarly, for modes $b_{1}$ and $b_{2}$.
The experimental apparatus of Fig. 2b realizes a double interferometer
operating with a single BS, avoiding the need for any active phase
stabilization.

The last term of the operator ${\cal O}$, $z_{1}z_{1}^{\prime }\cdot
x_{1}x_{1}^{\prime }\cdot z_{2}x_{2}^{\prime }\cdot x_{2}z_{2}^{\prime }$,
is measured by the same setup of Fig. 2b, simply by removing the half
waveplate ($HW^{\ast }$) which performs the $|\Phi ^{-}\rangle \rightarrow
|\Psi ^{-}\rangle $ transformation. This operation realizes the Bell state
analysis performed in the teleportation experiments of Ref. \cite{15} and
allows to evaluate the expectation values of the operator $%
z_{1}z_{1}^{\prime }\cdot x_{1}x_{1}^{\prime }\cdot z_{2}x_{2}^{\prime
}\cdot x_{2}z_{2}^{\prime }$ \cite{13}.\ In this measurement the Alice's and
Bob's detectors, $(D_{a1}-D_{a2})$ and $(D_{b1}-D_{b2})$ perform the
measurements in the $H-V$\ basis.

The whole experiment has been carried out by a sequence of measurements each
one lasting an average time of $30\sec $. The experimental results
corresponding to the measurement of the nine terms of ${\cal O}$ are
summarized in the histogram shown in Fig. 3. The experimental value of $%
{\cal O}$, obtained after summation over all the measured values, ${\cal O}%
=8.114\pm 0.011$, which, corresponding to a violation of the inequality by $%
101-\sigma $ standard deviations, demonstrates a large contradiction with
local realism.

In this letter we have experimentally demonstrated the nonlocal character of
two photon hyper-entangled states by an ''all versus nothing'' test of
quantum mechanics. In addition to the conceptual relevance of this result
concerning one of the most intriguing fundamental properties of Nature, the
content of the present work may be viewed as a clear demonstration of the
power and the flexibility of the new source here applied for the first time
to the generation of hyper-entangled states. The simplicity and reliability
of the optical scheme together with the conceptual relevance of the
underlying quantum process here realized, i.e. the effective doubling to the
extension of the Hilbert space spanned by the state of the generated
particles, is expected to be appreciated in the near future as a useful and
far reaching resource of QI technology. During the preparation of this work,
another ''all versus nothing'' experiment has been performed with a
different two photon hyperentanglement source \cite{21}.

Thanks are due to A. Cabello and F. Sciarrino for useful discussions. This
work was supported by the FET European Network on Quantum Information and
Communication (Contract IST-2000-29681: ATESIT), FIRB-MIUR 2001 and
PRA-INFM\ 2002 (CLON).

{\it Note added in proof:} In the present work, the adoption of the same
apparatuses for the measurement of couples of operators as and implies the
supplementary assumption that the numerical results of the measurement of $%
z_{1}z_{1}^{\prime }$ is equal to the product of the results of $z_{1}$ and $%
z_{1}^{\prime }$ measured separately. The same argument holds for the
measurement of the other operators: $x_{1}$ $x_{1}^{\prime }$ and $%
x_{1}x_{1}^{\prime }$, $z_{2}$ $x_{2}^{\prime }$ and $z_{2}x_{2}^{\prime }$, 
$x_{2}$ $z_{2}^{\prime }$ and $x_{2}$ $z_{2}^{\prime }$. We are presently
adopting a new, more complete measurement method which does not imply that
extra assumption. The preliminary results obtained with this new method for
the three operators, $z_{1}^{\prime }\cdot z_{2}^{\prime }=-0.9893\,\pm
\,0.0031$, $z_{1}\cdot z_{2}=-0.9348\,\pm \,0.0037$, and $z_{1}z_{1}^{\prime
}\cdot z_{2}\cdot z_{2}^{\prime }=0.9218\,\pm \,0.0037$, are in full
agreement with the data presented in this paper.

\centerline{\bf Figure Captions}

\vskip 8mm

\parindent=0pt

\parskip=3mm

\begin{description}
\item[Fig. 1 - ]  a) Parametric source of polarization-momentum
hyper-entangled two photon states. Phase setting $\theta =0,\pi $ and $%
\varphi =0,\pi $ are obtained by micrometric translation of the spherical
back mirror and by tilting of the glass plate on mode $a_{1}$; b) Spatial
coupling of input modes $a_{1}-b_{1}$, $a_{2}-b_{2}$ on the $BS$ plane. The $%
BS$ output modes, $a_{1}^{\prime }-b_{1}^{\prime }$, $a_{2}^{\prime
}-b_{2}^{\prime }$ are also shown. $M_{1}\,$and $M_{2}$ are mirrors inserted
to separate upper and lower modes; c) Coincidence rate $C(a_{1}^{\prime
},b_{1}^{\prime })$ vs. $\Delta x$: $\theta =0$, $\phi =0$ (upper curve), $%
\theta =0$, $\phi =\pi $\ (lower curve).

\item[Fig. 2 - ]  Experimental apparatus to measure the expectation values
of the nine operators compairing in ${\cal O}$. a) Modes $a_{1},\,b_{1}$, $%
a_{2}\ $and $b_{2},$ separated by the pick-off mirrors $M_{1}\,$and $M_{2},$
are directly coupled to detectors; polarization analysis is performed in the
orthonormal bases, either $H-V$ or $D-\bar{D}$, by rotating the halfwave
plates HW before the detectors. The halfwave plate $HW^{\ast }$ performing
the $\left| \Phi \right\rangle \rightarrow \left| \Psi \right\rangle $
transformation is also shown. By this configuration one can evaluate the
values of $z_{1}\cdot z_{2},$ $x_{1}\cdot x_{2},$ $z_{1}^{\prime }\cdot
z_{2}^{\prime },\,z_{1}z_{1}^{\prime }\cdot z_{2}\cdot z_{2}^{\prime }$, and 
$x_{1}\cdot z_{1}^{\prime }\cdot x_{2}z_{2}^{\prime }$. b) The two mode sets 
$a_{1}-b_{1}$ and $a_{2}-b_{2}$ are spatially combined onto the $BS$ before
being coupled to detectors. In this way one can perform the transformation
on the momentum\ basis, and measure the values of $x_{1}^{\prime }\cdot
x_{2}^{\prime }$, $x_{1}x_{1}^{\prime }\cdot x_{2}\cdot x_{2}^{\prime }$,
and $z_{1}\cdot x_{1}^{\prime }\cdot z_{2}x_{2}^{\prime }$. The same
apparatus is used for performing the Bell state analysis by removing $%
HW^{\ast }$ plate, in order to evaluate the value of $z_{1}z_{1}^{\prime
}\cdot x_{1}x_{1}^{\prime }\cdot z_{2}x_{2}^{\prime }\cdot
x_{2}z_{2}^{\prime }.$

\item[Fig. 3 - ]  Barchart of expectation values for the nine operators
involved in the experiment. The following results have been obtained: $%
z_{1}\cdot z_{2}=-0.9428\pm 0.0030,$ $z_{1}^{\prime }\cdot z_{2}^{\prime
}=-0.9953\pm 0.0033,$ $z_{1}z_{1}^{\prime }\cdot z_{2}\cdot z_{2}^{\prime
}=0.9424\pm 0.0030,$ $x_{1}\cdot x_{2}=-0.9215\pm 0.0033,$ $x_{1}\cdot
z_{1}^{\prime }\cdot x_{2}z_{2}^{\prime }=0.9217\pm 0.0033,$ $x_{1}^{\prime
}\cdot x_{2}^{\prime }=-0.8642\pm 0.0043,$ $z_{1}\cdot x_{1}^{\prime }\cdot
z_{2}x_{2}^{\prime }=0.8039\pm 0.0040,$ $x_{1}x_{1}^{\prime }\cdot
x_{2}\cdot x_{2}^{\prime }=0.8542\pm 0.0040,$ $z_{1}z_{1}^{\prime }\cdot
x_{1}x_{1}^{\prime }\cdot z_{2}x_{2}^{\prime }\cdot x_{2}z_{2}^{\prime
}=-0.8678\pm 0.0043.$
\end{description}


\begin{references}
\bibitem{1}  J. S. Bell {\it Physics,} {\bf 1}, 165 (1964); J. F. Clauser,
M. A. Horne, A. Shimony and R. A. Holt, {\it Phys. Rev. Lett. }23, 880
(1969); J. F. Clauser and M. A. Horne, Phys. Rev. D 10, 526 (1974).

\bibitem{2}  A.Einstein, B.Podolsky and N.Rosen, Phys. Rev. {\bf 47}, 777
(1935).

\bibitem{3}  S. J. Freedman, J. F. Clauser, {\it Phys. Rev. Lett.} {\bf 28},
938 (1972).

\bibitem{4}  A. Aspect, P. Grangier, G. Roger, {\it Phys. Rev. Lett.} {\bf 47%
}, 460 (1981); A. Aspect, P. Grangier, G. Roger, {\it Phys. Rev. Lett.} {\bf %
49}, 91 (1982); A. Aspect, J.Dalibard, G. Roger, {\it Phys. Rev. Lett.} {\bf %
49}, 1804 (1982).

\bibitem{5}  L. Hardy, {\it Phys. Rev. Lett.} {\bf 71}, 1665 (1993).

\bibitem{6}  D. Boschi, S. Branca, F. De Martini, L. Hardy, {\it Phys. Rev.
Lett. }{\bf 79}, 2755 (1997); M. Barbieri, F. De Martini, G. Di Nepi, P.
Mataloni, {\it Phys. Lett. A} {\bf 334}, 23 (2005).

\bibitem{7}  N.D. Mernin, {\it Phys. Today}, {\bf 43}(6),9 (1990).

\bibitem{8}  D. M. Greenberger, M. A. Horne, A. Zeilinger, {\it Am. J. Phys.}
{\bf 58}, 1131 (1990).

\bibitem{9}  J.W. Pan {\it et al.}, {\it Nature} (London){\bf 403}, 515
(2000).

\bibitem{10}  Z. Zhao {\it et al.}, {\it Phys. Rev. Lett.}{\bf 91}, 180401
(2003).

\bibitem{11}  A. Cabello, {\it Phys. Rev. Lett.}{\bf 86}, 1911 (2001); A.
Cabello, {\it Phys. Rev. Lett.}{\bf 87}, 010403 (2001).

\bibitem{12}  D. Collins, N. Gisin, N. Linden, S. Massar, and S. Popescu,
Phys. Rev. Lett. {\bf 88}, 040404 (2002); D. Kaszlikowski, L. C. Kwek, J-L
Chen, M. Zukowski, and C. H. Oh, Phys. Rev. A {\bf 65}, 032118 (2002).

\bibitem{13}  Z.B Chen, J.W. Pan, Y.D. Zhang, C. Brukner, A. Zeilinger, {\it %
Phys. Rev. Lett.}, {\bf 90}, 160408 (2003).

\bibitem{14}  P. G. Kwiat, {\it Journal of Modern Optics}, {\bf 44}, 2173
(1997); P. G. Kwiat and H. Weinfurter, {\it Phys. Rev. A},{\bf \ 58}, R2623
(1998). Two photon space-time and spin double entanglement has been
considered by the UMBC\ group: T. B. Pittman, Y. H. Shih, A. V. Sergienko,
and M. H. Rubin {\it Phys. Rev. A},{\bf \ 51}, 3495 (1995).

\bibitem{15}  D. Boschi, S. Branca, F. De Martini, L. Hardy and S. Popescu, 
{\it quant-ph/}9710013; {\it Phys.\ Rev.\ Lett.}\ {\bf 80}, 1121 (1998).

\bibitem{16}  S. Ghosh, G. Kar, A. Roy, A. Sen (De) and U. Sen, {\it Phys.\
Rev.\ Lett.}\ {\bf 87}, 277902 (2001); J. Calsamiglia, N. Lukenhaus, {\it %
Appl. Phys. B: Lasers Opt.}, {\bf 72}, 67 (1999).

\bibitem{17}  C. Cinelli, G. Di Nepi, F. De Martini, M. Barbieri, and P.
Mataloni, {\it Phys. Rev. A}, {\bf 70}, 022321 (2004).

\bibitem{18}  M.Barbieri, F.De Martini, G.Di Nepi, P.Mataloni, G.M.D'Ariano
and C.Macchiavello, {\it Phys.Rev.Lett }{\bf 91}{\it , }227901{\it \ }(2003)%
{\it ; }M. Barbieri, F. De Martini, G. Di Nepi, P. Mataloni, {\it Phys. Rev.
Lett.}, {\bf 92}, 177901 (2004).

\bibitem{19}  S. P. Walborn, S. Pad\`{u}a and C. H. Monken, {\it Phys. Rev.
A }{\bf 68}, 042313 (2003).

\bibitem{20}  G.Di Giuseppe, F. De Martini, D. Boschi and S. Branca, {\it %
Fortschr. Phys} {\bf 46}, 643 (1998).

\bibitem{21}  T. Yang $et$ $al.$, {\it Phys. Rev. Lett.}, {\bf 95}, 240406
(2005).
\end{references}
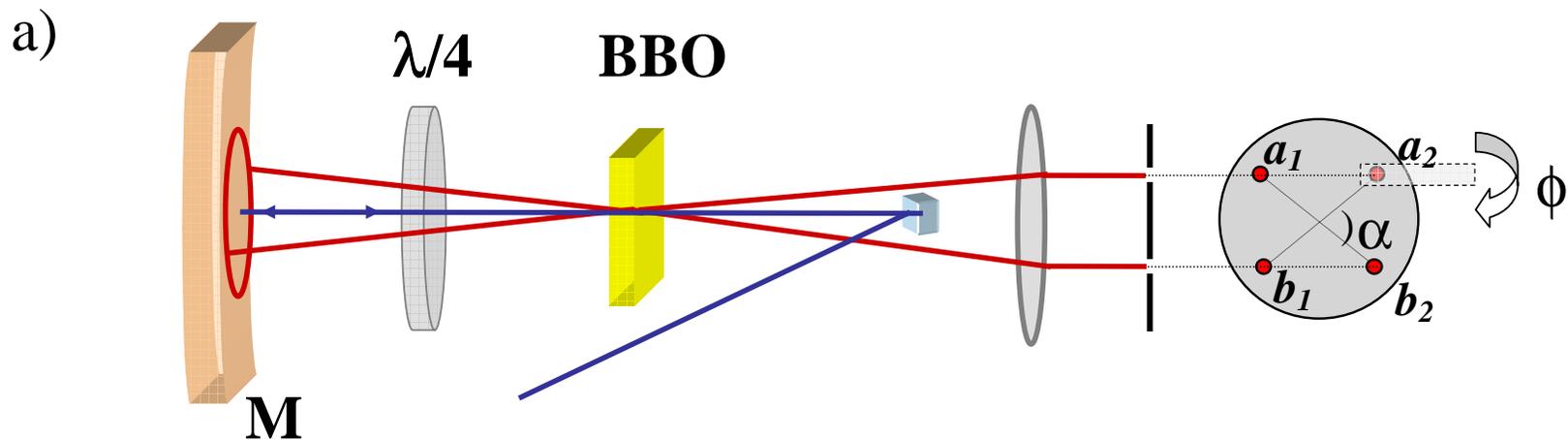
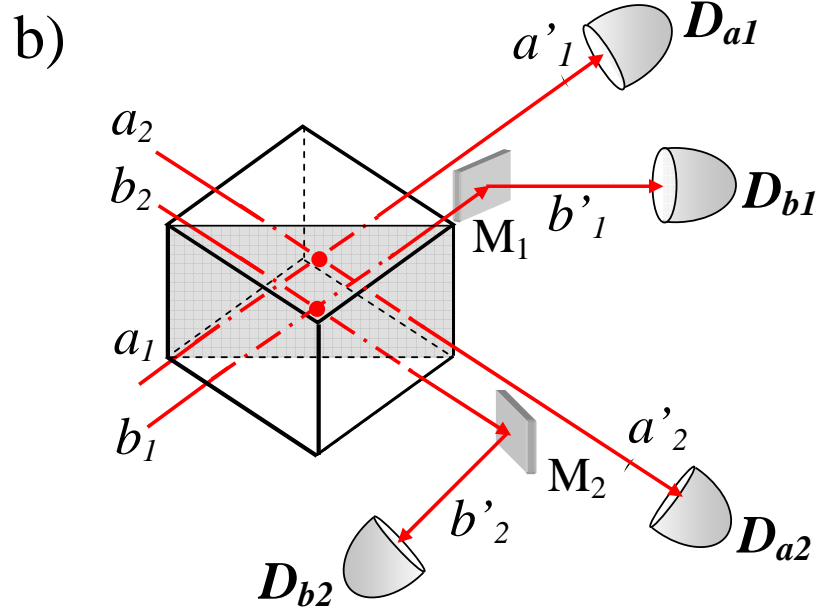
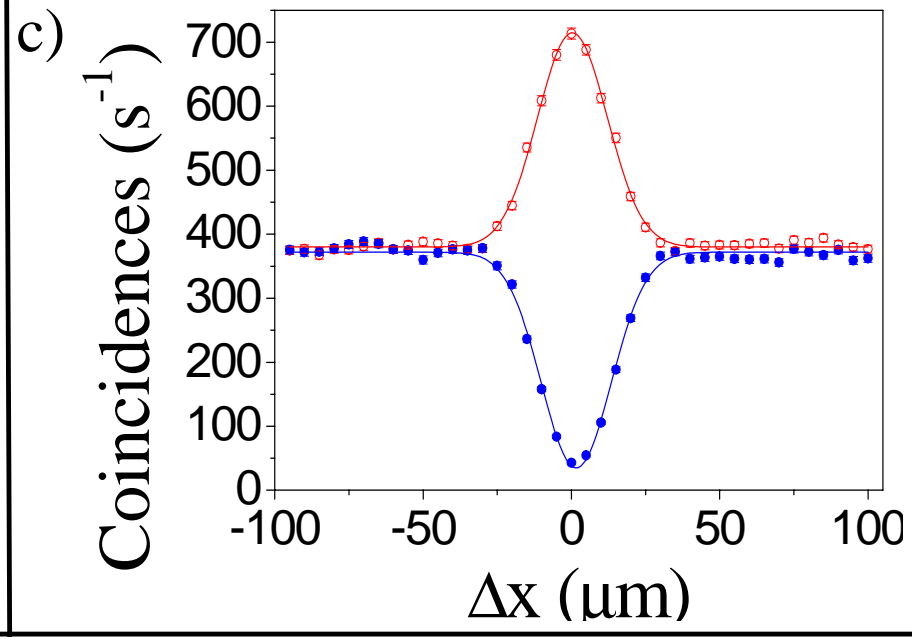
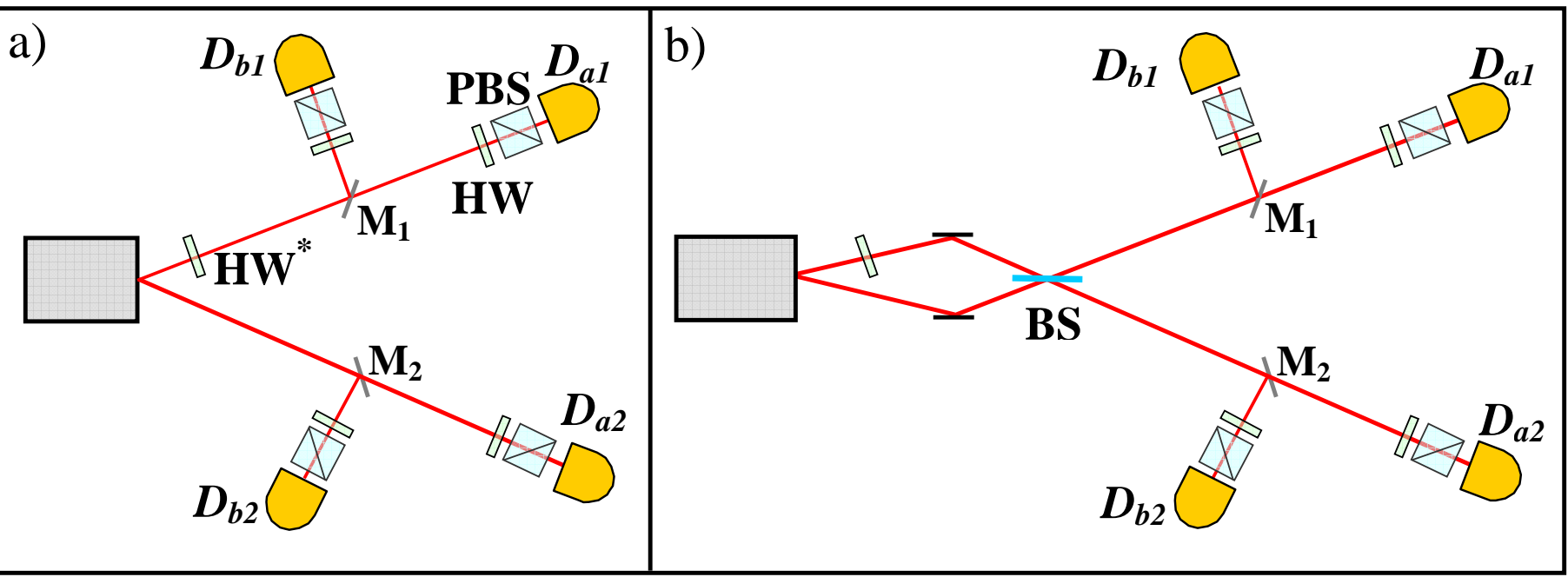
\end{document}